\documentclass[a4paper]{article}
\usepackage{hyperref}
\usepackage{multirow}

\usepackage{INTERSPEECH2020}

\title{Speaker Conditional WaveRNN: Towards Universal Neural Vocoder for Unseen Speaker and Recording Conditions}
\name{Dipjyoti Paul$^1$, Yannis Pantazis$^2$ and Yannis Stylianou$^1$}
\address{
  $^1$Computer Science Department, University of Crete\\
  $^2$Inst. of Applied and Computational Mathematics, Foundation for Research and Technology - Hellas}
\email{dipjyotipaul@csd.uoc.gr, pantazis@iacm.forth.gr, yannis@csd.uoc.gr}

\begin{document}

\maketitle
\vspace{-2mm}
\begin{abstract}
Recent advancements in deep learning led to human-level performance in single-speaker speech synthesis. However, there are still limitations in terms of speech quality when generalizing those systems into multiple-speaker models especially for unseen speakers and unseen recording qualities. For instance, conventional neural vocoders are adjusted to the training speaker and have poor generalization capabilities to unseen speakers. In this work, we propose a variant of WaveRNN, referred to as speaker conditional WaveRNN (SC-WaveRNN). We target towards the development of an efficient universal vocoder even for unseen speakers and recording conditions. In contrast to standard WaveRNN, SC-WaveRNN exploits additional information given in the form of speaker embeddings. Using publicly-available data for training, SC-WaveRNN achieves significantly better performance over baseline WaveRNN on both subjective and objective metrics. In MOS, SC-WaveRNN achieves an improvement of about 23\% for seen speaker and seen recording condition and up to 95\% for unseen speaker and unseen condition. Finally, we extend our work by implementing a multi-speaker text-to-speech (TTS) synthesis similar to zero-shot speaker adaptation.
In terms of performance, our system has been preferred over the baseline TTS system by 60\% over 15.5\% and by 60.9\% over 32.6\%, for seen and unseen speakers, respectively.
\end{abstract}
\section{Introduction}

Speech synthesis has received attention in the research community as voice interaction systems have been implemented in various applications, such as personalized Text-to-Speech (TTS) systems, voice conversion, dialogue systems and navigations \cite{dutoit1997introduction, taylor2009text, stylianou1998continuous, paul2019non}.
In the past, conventional statistical parametric speech synthesis (SPSS) exhibited high naturalness under best-case conditions \cite{zen2009statistical, king2011introduction}. Hybrid synthesis was also proposed as a way to take advantage of both SPSS and unit-selection approach \cite{qian2012unified,merritt2016deep}. Most of these TTS systems consist of two modules: the first module converts textual information into acoustic features while the second one, i.e., the vocoder, generates speech samples from the previously generated acoustic information.

Traditional vocoder approaches mostly involved source-filter model for the generation of speech parameters  \cite{mcaulay1986speech, moulines1990pitch, kawahara1999restructuring, morise2016world}. The parameters were defined by voicing decisions, fundamental frequency (F0), spectral envelope or band aperiodicities. Algorithms like Griffin-Lim utilized spectral representation to generate speech \cite{griffin1984signal,perraudin2013fast}. However, the speech quality of such vocoders was restricted by the inaccuracies in parameter estimation. Recently, the naturalness of vocoders has been significantly improved by benefiting from direct waveform modeling approach. Neural vocoders like WaveNet utilize a autoregressive generative model that can reconstruct waveform from intermediate acoustic features \cite{oord2016wavenet, tamamori2017speaker}. To overcome the time complexity at inference, parallel wave generation approach was adopted to generate speech in real time \cite{oord2017parallel, ping2018clarinet}. Wave Recurrent Neural Networks (WaveRNN) which employs recurrent layers increases the efficiency of sampling without compromising their quality \cite{kalchbrenner2018efficient}. In particular, it can realize real-time high-quality synthesis by introducing a gated recurrent unit (GRU). Although, WaveRNN has been suggested focusing on text-to-speech synthesis, our work exercises it as a vocoder while changing the conditioning criteria from linguistic information to acoustic information. Other recent works have been also found in literature, notable among them are SampleRNN \cite{mehri2016samplernn}, WaveGlow \cite{prenger2019waveglow}, LPCNet \cite{valin2019lpcnet} and MelNet \cite{vasquez2019melnet}.

Techniques in neural vocoders involve data-driven learning and are prone to specialize to the training data which leads to poor generalization capabilities. Moreover, in multi-speaker scenarios, it is practically impossible to cover all possible in-domain (or seen) and out-of-domain (or unseen) cases in the training database. Previous studies also attempted to improve adaptation capabilities of vocoders \cite{sisman2018voice}, either with or without providing speaker information \cite{liu2018wavenet, hayashi2017investigation}. However, these studies did not address the generalization capabilities for unseen out-of-domain data. In \cite{lorenzo2018towards}, a potential universal vocoder was introduced claiming that speaker encoding is not essential to train a high-quality neural vocoder.

\begin{figure*}[t!]
  \begin{center}
    \includegraphics[width=6.6in]{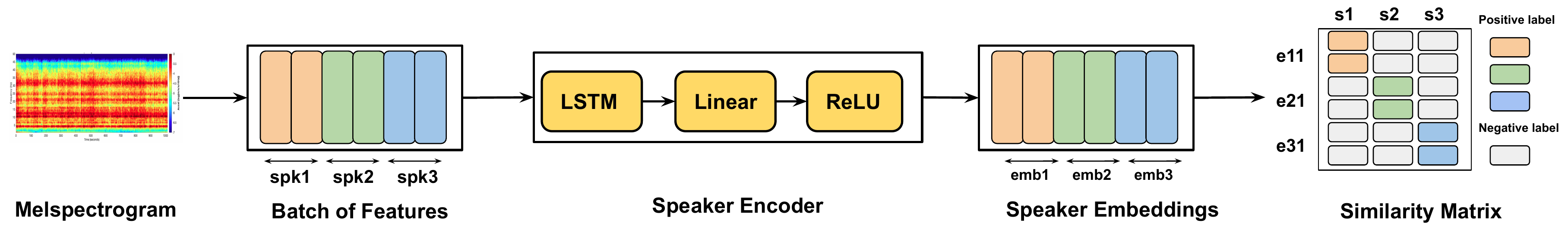}
  \end{center}
  \vspace{-7mm}
  \caption{System overview of speaker encoder \cite{wan2018generalized}. Features, speaker embeddings and similarity scores from different speakers are represented by different color codes. 'spk' denotes speakers and 'emb' represents embedding vectors.}
  \vspace{-6mm}
  \label{spk-encode}
\end{figure*}

Inspired by the performance and computational aspects of WaveRNN, we propose a novel approach for designing a universal WaveRNN vocoder. The proposed universal vocoder-speaker conditional WaveRNN (SC-WaveRNN) explores the effectiveness of explicit speaker information, i.e., speaker embeddings as a condition and improves the quality of generated speech across broadest possible range of speakers without any adaptation or retraining. Even though conventional WaveRNN is capable of modeling good temporal structure for a single speaker, it fails to capture the dynamics of multiple speakers. We have experimentally demonstrated that our proposed SC-WaveRNN overcomes such limitation by modeling temporal structure from a large variability of data, making it possible to generate high-quality synthetic voices. Our work involves independent training of a speaker-discriminative neural encoder on a speaker verification (SV) task using a state-of-the-art generalized end-to-end loss \cite{wan2018generalized}. The SV model, trained on a large amount of disjoint data, can attain robust speaker representations that are independent of channel conditions and captures large space of speaker characteristics. Coupling such speaker information with the speech synthesis training also reduces the need to obtain ample high-quality multi-speaker training data. At the same time, it increases the model's ability to generalize. Experimental results based on both objective and subjective evaluation confirms that the proposed method achieves better speaker similarity and perceptual speech quality than baseline WaveRNN in both seen and unseen speakers.

In parallel with the above-mentioned studies on universal vocoder, there has been substantial development in multi-speaker TTS where speaker encoder is jointly trained with TTS \cite{chen2018sample, park2019multi}. These jointly-trained speaker encoders lead to poor inference performance when applied on data which are not included in the training dataset. Fine-tuning pretrained TTS model in combination with speaker embeddings was addressed in \cite{deng2018modeling, hu2019neural, arik2018neural}. Such approaches always require transcribed adaptation data along with more computational time and resources to adapt to a new speaker. To overcome this, TTS models can be adapted from a few seconds of target speaker’s voice in a zero-shot manner by solely using speaker embedding without retraining the entire model. \cite{jia2018transfer,chen2019cross,cooper2019zero}.

Unfortunately, limitations still exist and human-level naturalness is not achieved yet. Additionally, prosody information was mismatched especially for unseen speakers. To address those issues, we first train a multi-speaker Tacotron which is conditioned on the speaker embeddings obtained from the independently-trained speaker encoder. Tacotron \cite{wang2017tacotron} is a sequence-to-sequence network which predicts mel-spectrograms from text. Next, we incorporate the proposed SC-WaveRNN as a vocoder using the same speaker encoder and synthesize the temporal waveform from the sequence of Tacotron's mel-spectrograms. We compare our system with the baseline TTS method \cite{cooper2019zero} which studies the effectiveness of several neural speaker embeddings in the context of zero-shot TTS. Our results demonstrate that the proposed zero-shot TTS system outperforms baseline zero-shot TTS in \cite{cooper2019zero} in-terms of both speech quality and speaker similarity on both seen and unseen conditions.


\vspace{-2mm}
\section{Neural Speaker Encoder}

\begin{figure}[h!]
  \begin{center}
  \vspace{-2mm}
    \includegraphics[height=1.3in,width=1.7in]{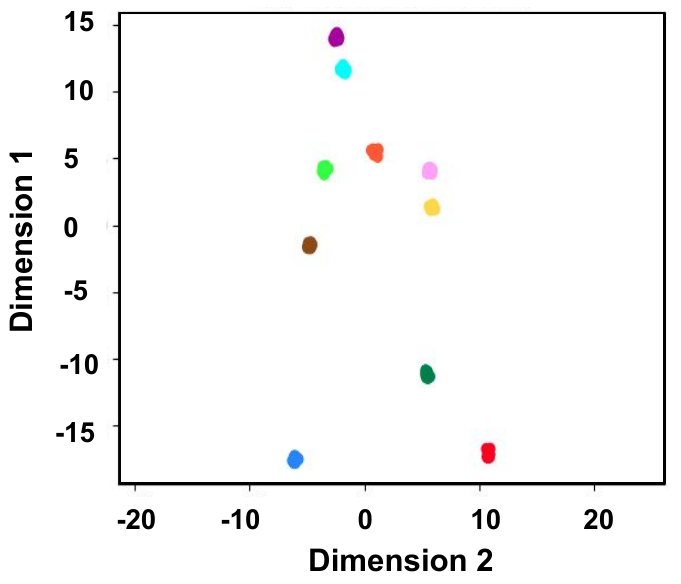}
  \end{center}
  \vspace{-6mm}
  \caption{\small UMAP projection of 10 utterances for each of the 10 speakers. Different colors represent different speakers.}
  \vspace{-6mm}
  \label{umap}
\end{figure}

Our work highlights the importance of speaker encoder in universal vocoders through the application of generalized end-to-end (GE2E) SV task trained on thousands of speakers \cite{wan2018generalized}.
The encoder network initially computes frame-level feature representation and then summarizes them to utterance-level fixed-dimensional speaker embeddings. Next, the classifier operates on GE2E loss, where embeddings from the same speaker have high cosine similarity and embeddings from different speakers are far apart in the embedding space. As depicted in Fig. \ref{umap}, Uniform Manifold Approximation and Projection (UMAP) shows that the speaker embeddings are perfectly separated with large inter-speaker distances and very small intra-speaker variance.
\vspace{-1mm}
\subsection{Training Encoder Network}
Speaker encoder structure is depicted in Figure \ref{spk-encode}. The log mel-spectrograms are extracted from speech utterance of arbitrary window length. The feature vectors are then assembled in the form of a batch that contains $S$ different speakers, and each speaker has $U$ utterances. Each feature vector $\mathbf{x}_{ij}$ ($1 \leq i \leq S$ and $1\leq j \leq U$) represents the features extracted from speaker $i$ utterance $j$. The features $\mathbf{x}_{ij}$ are then passed to an encoder architecture. The final embedding vector $\mathbf{e}_{ij}$ is L2 normalized and they are calculated by averaging on each window separately.

\subsection{Generalized End-to-End Loss}
During training, embedding of all utterance for a particular speaker should be closer to the centroid of that particular speaker’s embeddings, while far from other speakers’ centroids. The similarity matrix $\mathbf{SM}_{ij,k}$ is defined as the scaled cosine similarities between each embedding vector $\mathbf{e}_{ij}$ to all speaker centroids $\mathbf{c}_{k}$ ($1 \leq i,k \leq S$ and $1\leq j \leq U$).
\vspace{-2mm}
\[\vspace{-2mm}
    \mathbf{SM}_{ij,k} =
\begin{cases}
    w \cdot cos(\mathbf{e}_{ij},\mathbf{c}_{i}^{-j}) + b  & \text{if } k = i\\
    w \cdot cos(\mathbf{e}_{ij},\mathbf{c}_{k}) + b   & \text{otherwise}
\end{cases}
\]
\[\vspace{-1.5mm}
    where \: \: \mathbf{c}_{i}^{-j} = \frac{1}{U-1}\displaystyle\sum_{u=1;u \neq j}^U \mathbf{e}_{iu} \: and \: \mathbf{c}_{k} = \frac{1}{U}\displaystyle\sum_{u=1}^U \mathbf{e}_{ku}
\]

Here, $w$ and $b$ are trainable parameter. The ultimate GE2E loss $L$ is the accumulative loss over similarity matrix ($1 \leq i \leq S$ and $1\leq j \leq U$) on each embedding vector $\mathbf{e}_{ij}$:
\vspace{-4mm}
\[\vspace{-1.7mm}
    L(\mathbf{x};\mathbf{w}) = \displaystyle\sum_{i,j} L(\mathbf{e}_{ij}) = - \mathbf{SM}_{ij,i} + \log \displaystyle\sum_{k=1}^{S} \exp(\mathbf{SM}_{ij,k})
\]
The use of softmax function on similarity matrix makes the output equals to 1 iff $k = i$, otherwise the output is 0.

\vspace{-2mm}
\section{Speaker conditional WaveRNN}

In literature, convolutional models have been thoroughly explored and achieved excellent performance in speech synthesis \cite{oord2016wavenet, ping2018clarinet} yet they are prone to instabilities. Recurrent neural network (RNN) is expected to provide a more stable high-quality speech due to the persistence of the hidden state.
\vspace{-2mm}
\subsection{Preliminaries}
\begin{figure}[h!]
  \begin{center}
  \vspace{-2mm}
    \includegraphics[height=2.4in,width=3.25in]{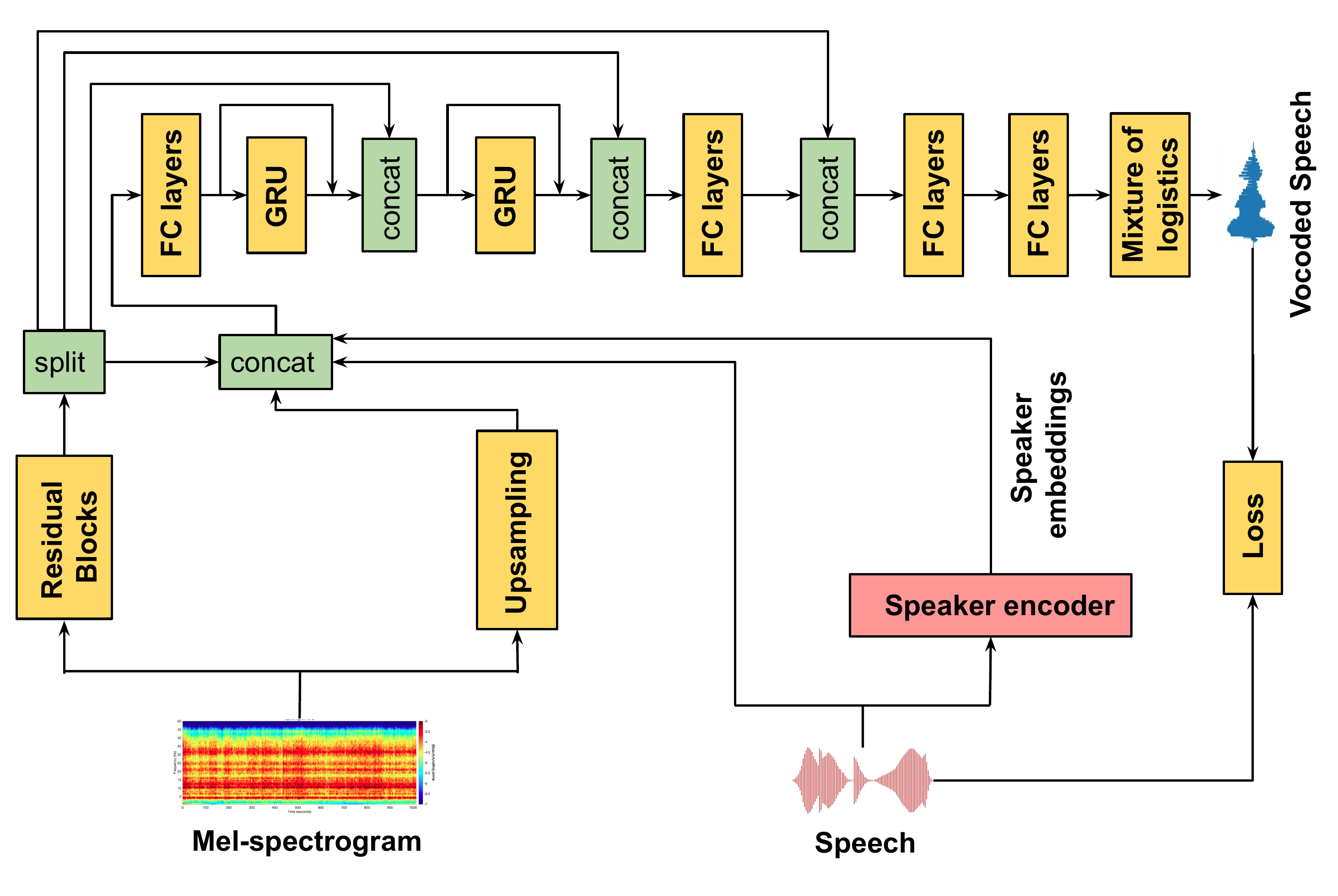}
  \end{center}
  \vspace{-8mm}
  \caption{\small Block diagram of proposed SC-WaveRNN training.}
  \vspace{-7mm}
  \label{wavernn}
\end{figure}

Our WaveRNN implementation is based on the repository\footnote{\url{https://github.com/fatchord/WaveRNN}} which is heavily inspired by WaveRNN training \cite{kalchbrenner2018efficient}.  This architecture is a combination of residual blocks and upsampling network, followed by GRU and FC layers as depicted in Fig. \ref{wavernn}. 
The architecture can be divided into two major networks: conditional network and recurrent network. The conditioning network consists of a pair of residual network and upsampling network with three scaling factors. At the input, we first map the acoustic features i.e., mel-spectrograms to a latent representation with the help of multiple residual blocks. The latent representation is then split into four parts which will later be fed as input to the recurrent network. The upsampling network is implemented to match the desired temporal size of input signal. The outputs of these two convolutional networks i.e., residual and upsampling networks along with speech are fed into the recurrent network. As part of the recurrent network, two uni-directional GRUs are employed with a few fully-connected (FC) layers at the end. By design, the overhead complexity is reduced  with less parameters and takes advantage of temporal context for better prediction.
\vspace{-2mm}

\subsection{Training WaveRNN with Speaker Embeddings}

The above auto-regressive model can generate state-of-the-art natural sounding speech, however, it needs large amounts of training data to train a stable high-quality model and scarcity of data remains a core issue. Moreover, a key challenge is its generalization ability. We observe degradation in speech quality and speaker similarity when the model generates waveforms from speakers that are not seen during training.

In order to assist the development of a stable universal vocoder and remove data dependency, we propose in this paper an alternative training module referred to as speaker conditional WaveRNN (SC-WaveRNN). In SC-WaveRNN, the output of the speaker encoder is used as additional information to control the speaker characteristics during both training and inference. The additional information plays a pivotal role in generating more stable high-quality speech across all speaker conditions. The direct estimation of raw audio waveform $ \mathbf{y} = \{y_{1}, y_{1} ,\cdots, y_{N}\}$ is described by the conditional probability distribution:
\vspace{-2mm}
\[
\vspace{-2mm}
    sc{\text -}wavernn(\mathbf{y}) = p(y_{t}|y_{t-1};\mathbf{h}_{t};\mathbf{e};\mathbf{\lambda})
\]
where $\mathbf{e}$ is the 256 dimension speaker embeddings vector. The speaker encoder is independently trained using large diversity of multi-speaker data that can generalize sufficiently to produce meaningful embeddings. The embedding vector $\mathbf{e}$ is computed in a utterance-wise manner. For each utterance, the final embedding vector is averaged over all frames and hence it is fixed for any utterance. The embedding vector is concatenated with the conditional network output and speech samples to form the conditional network. The details of the SC-WaveRNN algorithm is presented in Figure \ref{wavernn}. In addition, we apply continuous univariate distribution constituting a mixture of logistic distributions \cite{oord2017parallel} which allows us to easily calculate the probability on the observed discretized value $y$. Finally, discretized mix logistic loss is applied on the discretized speech.

\vspace{-2mm}
\section{Zero-shot Text-to-Speech}

The use of the auxiliary speaker encoder enables us to propose a TTS system capable of generating high-fidelity synthetic voice for unseen speakers without retraining Tacotron and vocoder model. Such speaker adaptation to completely new speakers is called zero-shot learning. This speaker-aware TTS system mimics voice characteristics from a completely unseen speaker with only a few seconds of speech sample.

\begin{figure}[h!]
  \begin{center}
  \vspace{-4mm}
    \includegraphics[height=1in,width=3.3in]{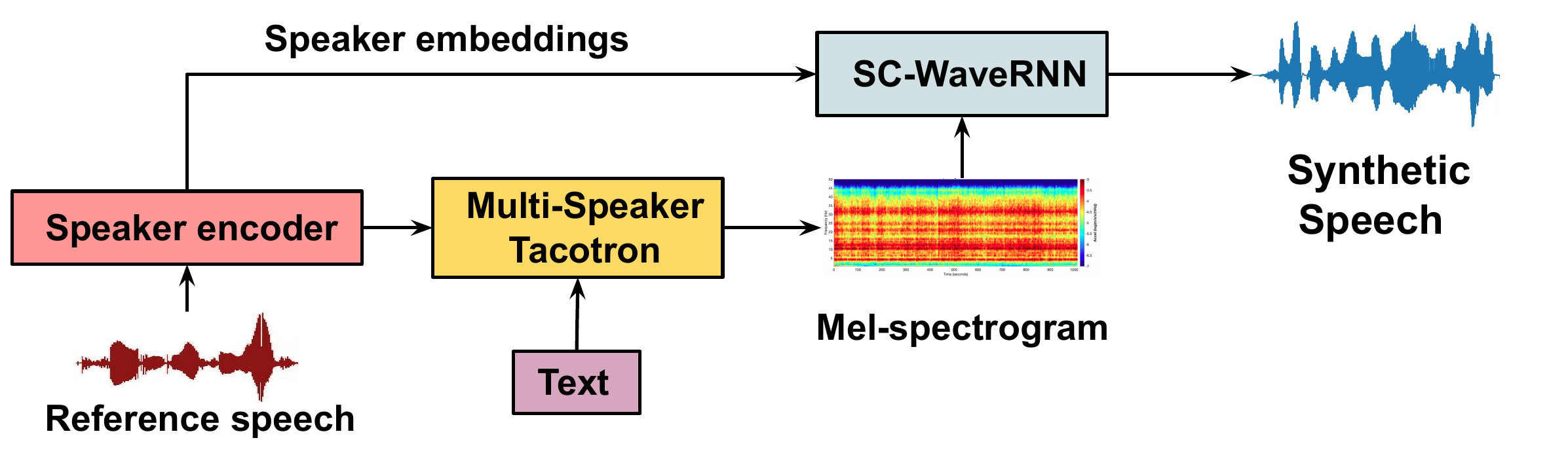}
    \vspace{-10mm}
  \end{center}
  \caption{\small Block diagram of the proposed zero-shot TTS.}
  \vspace{-3mm}
  \label{tts}
\end{figure}

Our proposed system is composed of three separately trained networks, illustrated in Figure \ref{tts}: (a) a neural speaker encoder, based on GE2E training, (b) a multi-speaker Tacotron  architecture \cite{wang2017tacotron}, which predicts a mel-spectrogram from text, conditioned on speaker embedding vector, and (c) the proposed speaker conditional WaveRNN, which converts the spectrogram into time domain waveforms. First, the speaker embeddings are extracted from each target speakers' utterance using the speaker encoder. At each time step, the embedding vector for the target speaker is then concatenated with the embeddings of the characters before fed into encoder-decoder module. The final output is mel-spectrograms. To convert the predicted mel-spectrograms into audio, we use SC-WaveRNN which is independently trained by conditioning on the additional speaker embeddings. Due to generalization capabilities of the models, combining multi-speaker Tacotron with SC-WaveRNN can achieve efficient zero-shot adaptation for unseen speakers. We compare the proposed zero-shot system with a recently proposed zero-shot TTS \cite{cooper2019zero} as baseline system. There, the best performing system uses multi-speaker Tacotron with gender-dependent WaveNet vocoders as TTS system and x-vector with learnable dictionary
encoding as speaker encoder network.

\vspace{-3mm}
\section{Experimental Setup}
The speaker encoder training has been conducted on three public dataset: LibriSpeech, VoxCeleb1 and VoxCeleb2 containing utterances from over 8k speakers \cite{jia2018transfer}. The log mel-spectrograms are first extracted from audio frames of width 25ms and step 10ms. Voice Activity Detection (VAD) and a
sliding window approach is used. The GE2E model consists of 3 LSTM layers of 768 cells followed by a projection to 256 dimensions. While training, each batch contains S = 64 speakers and U = 10 utterances per speaker.

Tacotron and WaveRNN models are trained using VCTK English corpus \cite{christophe2016cstr} from 109 different speakers. To evaluate generalization performance, we consider three scenarios: seen speakers-seen sound quality (SS-SSQ), unseen speakers-seen sound quality (UNS-SSQ) and unseen speakers-unseen sound quality (UNS-USQ). Seen speakers refers to the speakers that are already present in the training and unseen speakers are the new speakers during testing. Sound quality refers to the recording condition such as recording equipment, reverberation etc. We train the network using 100 speakers leaving 9 speakers for UNS-SSQ scenarios that are chosen to be a mix of genders and having enough unique utterances per speaker. CMU-ARCTIC database \cite{kominek2004cmu} is used for UNS-USQ scenario having 2 male and 2 female speakers. Moreover, to overcome the limited linguistic variability in VCTK data, we initially train Tacotron model on LJSpeech database as a ``warm-start'' training approach similar to \cite{cooper2019zero}. Code and sound samples can be found in \footnote{\url{https://dipjyoti92.github.io/SC-WaveRNN/}}.
\vspace{-2mm}
\section{Results and Discussion}
\subsection{Universal vocoder}
In this section, we evaluate the performance of vocoded speech shown in Table \ref{pesq-stoi}. To assess the effectiveness of speaker embeddings in SC-WaveRNN, PESQ and STOI objective measures are computed from 50 random samples. We carry out evaluations on three conditions: SS-SSQ, UNS-SSQ and UNS-USQ. The purpose of each condition is to evaluate the proposed vocoder not only on seen or unseen speakers but also for the quality of the recordings. As expected, seen scenarios perform better with respect to unseen samples. However, we observe that SC-WaveRNN significantly improves both the objective scores when compared to baseline WaveRNN for all scenarios.

\begin{table}[ht!]
\vspace{-3mm}
\footnotesize{}
\renewcommand{\arraystretch}{1.1}
\setlength{\tabcolsep}{3.4pt}
\caption{\small Objective evaluation tests.}
\centering
\vspace{-3mm}
\begin{tabular}{ccccccc}
\specialrule{1.25pt}{1pt}{1pt}
Methods         & \multicolumn{2}{c}{SS-SSQ} & \multicolumn{2}{c}{UNS-SSQ}& \multicolumn{2}{c}{UNS-USQ} \\ \cline{2-7}
         & PESQ                       & STOI                      & PESQ                        & STOI                       & PESQ                         & STOI                        \\ \specialrule{1.25pt}{1pt}{1pt}
WaveRNN & 2.2575                     & 0.8173                    & 2.1497                      & 0.7586                     & 1.4850                       & 0.8620                      \\
SC-WaveRNN & \textbf{2.7948}            & \textbf{0.9049}           & \textbf{2.8657}             & \textbf{0.8984}            & \textbf{1.8063}              & \textbf{0.9195}             \\ \specialrule{1.25pt}{1pt}{1pt}
\end{tabular}
\label{pesq-stoi}
\vspace{-3mm}
\end{table}
Concerning the perceptual assessment of speech quality and speaker similarity, two separate listening tests are reported: mean opinion score (MOS) and 'ABX' preference test. The subjects are asked to rate the naturalness of generated utterances on a scale of five-point (1:Bad, 2:Poor, 3:Fair, 4:Good, 5:Excellent). In the ’ABX’ test, experimental subjects have to decide whether a given reference sentence ’X’ is closer in speaker identity to one of ’A’ and ’B’ sentences, which are samples obtained either from the proposed or the baseline method, not necessarily in that order. Fifteen native and non-native English listeners participated in our listening tests. The evaluation results of both MOS and 'ABX' tests are demonstrated in Figure \ref{vocoder_subjective}. Error bars represent 95\% confidence intervals. For all seen and unseen scenarios, the MOS scores for the proposed SC-WaveRNN are much higher than the baseline WaveRNN (between 14\% to 95\% relative improvement). Under the same sound quality conditions (SS-SSQ and UNS-SSQ), although, the proposed technique is preferred in terms of speaker similarity preference test, a majority of preference is given to ’same preference’ option which indicates similar speaker characteristics for both methods. In contrast, experimental analysis shows a significant preference score (92\%) in unseen sound quality for proposed SC-WaveRNN. We conclude that additional speaker information in the form of embeddings is effective for improvements in naturalness and speaker similarity especially for unseen data and capable of achieving a truly universal vocoder. This is attributed by the fact that unseen scenarios are handled more efficiently by the model since additional embeddings are able to capture broad spectrum of speaker characteristics. Moreover, SC-WaveRNN does not compromise the performance in seen conditions.

\begin{figure}[t!]
  \begin{center}
    \includegraphics[height=0.8in, width=3.4in]{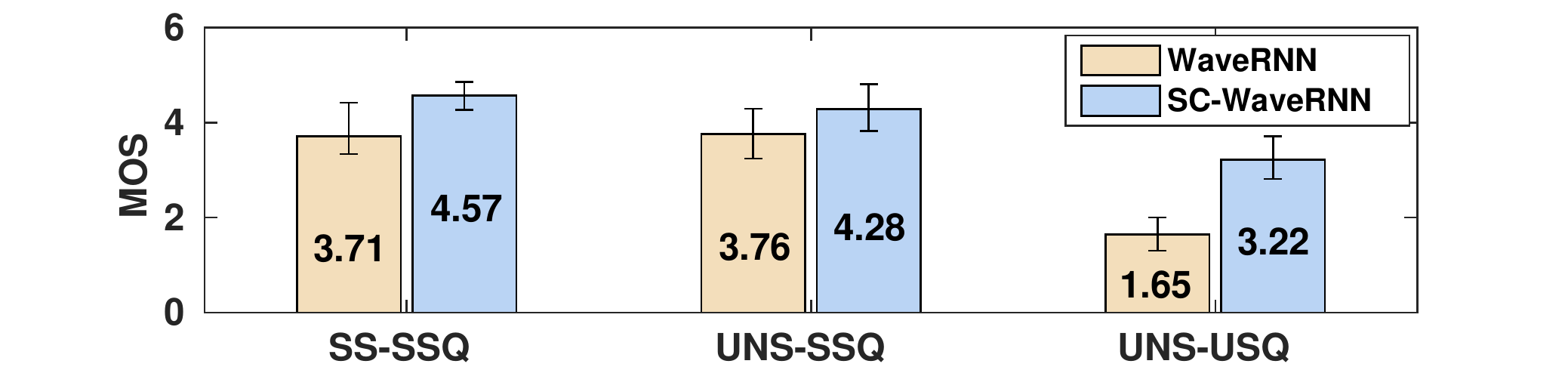}
    \vspace{-4mm}
  \end{center}
  \vspace{-4mm}
  \label{vocoder_subjective}
\end{figure}

\begin{figure}[t!]
  \begin{center}
  \vspace{-3mm}
    \includegraphics[height=0.9in, width=3.4in]{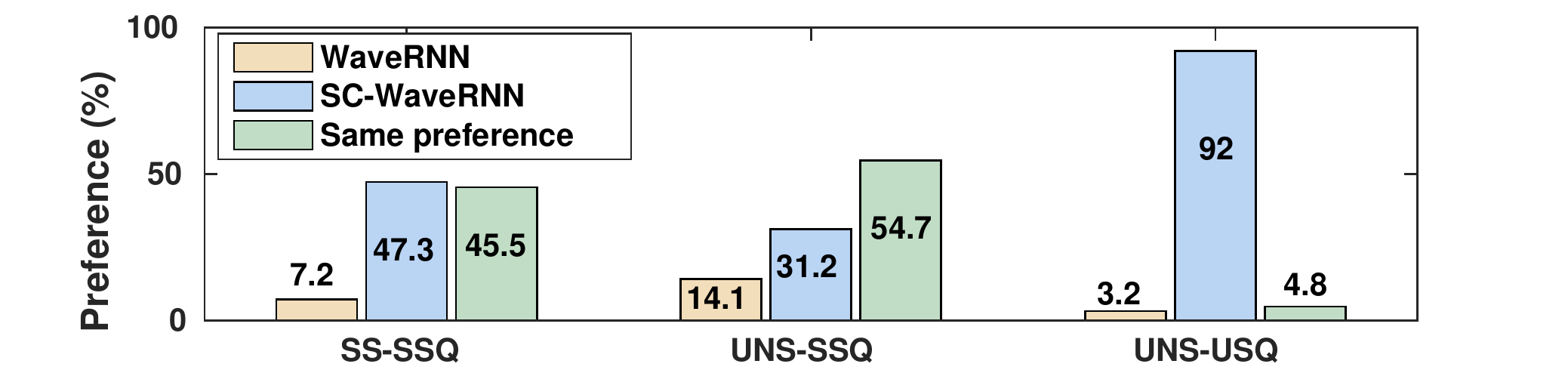}
    \vspace{-10mm}
  \end{center}
  \caption{\small Vocoder Subjective listening test (MOS) for speech quality and preference test in (\%) for speaker similarity.}
  \vspace{-5mm}
  \label{vocoder_subjective}
\end{figure}

\vspace{-2mm}
\subsection{Zero-shot TTS Synthesis}
To evaluate the performance of the proposed zero-shot TTS, MOS and 'ABX' test are employed, as depicted in Figure \ref{tts_subjective}. We subjectively evaluate both baseline \cite{cooper2019zero} and our methods by synthesizing sample utterances from seen speakers and unseen speakers. Different sound qualities are not considered in the evaluation experiments of zero-shot TTS. As expected, a gap between seen and unseen speakers are visible: seen speakers’ synthetic speech has slightly higher quality to unseen speakers. MOS scores indicate that proposed TTS is superior in quality with 19.2\% and 14.5\% relative improvement for seen and unseen speakers respectively. We also found that our proposed TTS mimic better speaker characteristics and shows significant improvement under both conditions. With regard to speaker similarity, the proposed TTS obtains the majority of preferences with 60\% and 60.9\% compared to 15.5\% and 32.6\% of the baseline TTS for seen and unseen speakers, respectively.
\begin{figure}[h!]
  \begin{center}
  \vspace{-3mm}
    \includegraphics[height=0.8in, width=3.3in]{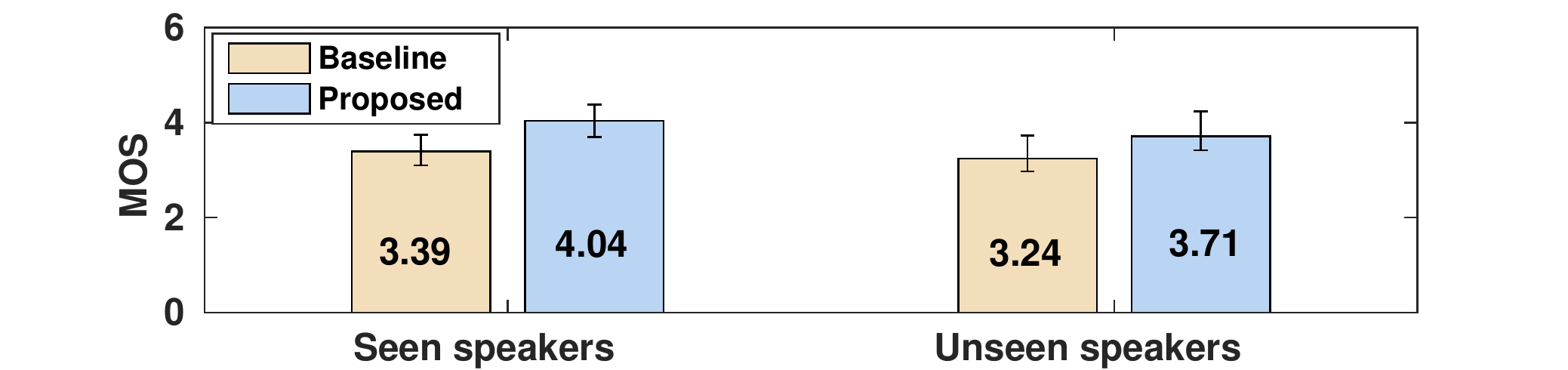}
    \vspace{-5mm}
  \end{center}
  \vspace{-8mm}
  \label{tts_subjective}
\end{figure}

\begin{figure}[h!]
  \begin{center}
  \vspace{-2mm}
    \includegraphics[height=0.9in, width=3.36in]{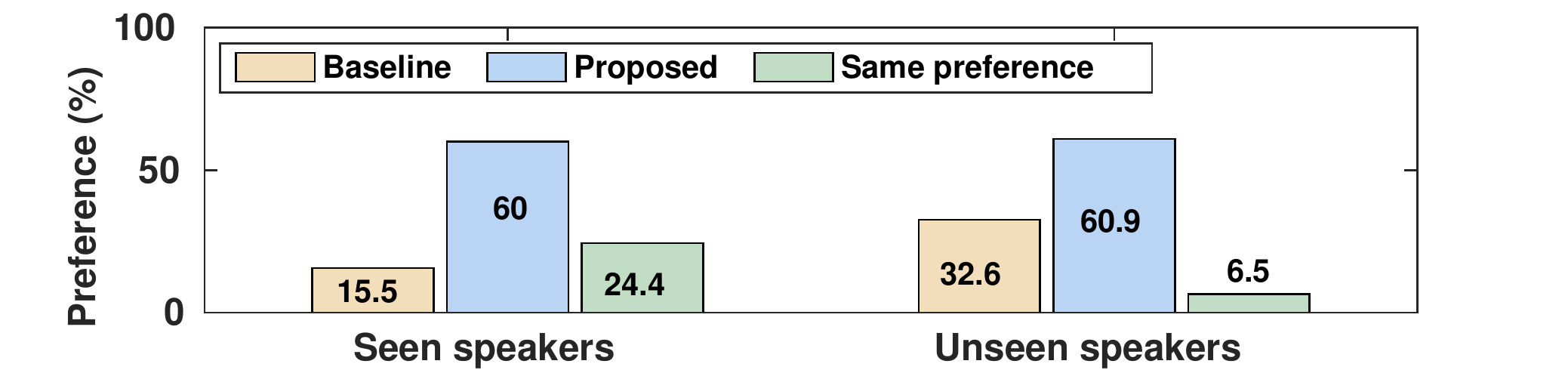}
    \vspace{-10mm}
  \end{center}
  \caption{\small Zero-shot TTS Subjective listening test (MOS) for speech quality and preference test for (\%) for speaker similarity.}
  \vspace{-5mm}
  \label{tts_subjective}
\end{figure}

\vspace{-2mm}
\section{Conclusions}
\vspace{-1mm}
In this paper, we proposed a robust universal SC-WaveRNN vocoder that is capable of synthesizing high-quality speech. The system was conditioned on extracted speaker embeddings which cover a very diverse range of seen and unseen conditions. The main advantage of SC-WaveRNN is its high controllability, since it improves multi-speaker vocoder training along with better generalization ability by allowing reliable transfer to unseen speaker characteristics. Furthermore, speaker conditioning is typically more data efficient and computationally less expensive than training separate models for each speaker. Subjective and objective evaluation revealed that the proposed method generated higher sound quality and speaker similarity than the baseline method. In addition, we extended our approach in devising an efficient zero-shot TTS system. We demonstrated that the proposed zero-shot TTS with universal vocoder can improve speaker similarity and naturalness of synthetic speech for seen and unseen speakers. In future, we list more experimentation on speaker embeddings and its effectiveness with unseen data.

\footnotesize{\bf Acknowledgements:} The work has received funding from the EUs H2020 research and innovation programme under the MSCA GA 67532 (the ENRICH network: www.enrich-etn.eu).

\clearpage
\bibliographystyle{IEEEtran}
\bibliography{template}

\end{document}